\documentclass[10pt,twocolumn,letterpaper]{article}
\usepackage[pagenumbers]{cvpr} 
\pdfoutput=1

\usepackage{graphicx}
\usepackage{amsmath}
\usepackage{amssymb}
\usepackage{booktabs}
\usepackage{multirow}
\usepackage{colortbl}

\usepackage[pagebackref,breaklinks,colorlinks]{hyperref}

\usepackage[capitalize]{cleveref}
\crefname{section}{Sec.}{Secs.}
\Crefname{section}{Section}{Sections}
\Crefname{table}{Table}{Tables}
\crefname{table}{Tab.}{Tabs.}

\def\cvprPaperID{8309
} 
\def\confName{CVPR}
\def\confYear{2023}

\begin{document}

\title{Role of Transients in Two-Bounce Non-Line-of-Sight Imaging}

\newcommand{\superscript}[1]{\ensuremath{^{\textrm{#1}}}}
\author{
    Siddharth Somasundaram\superscript{1} \hspace{1mm} 
    Akshat Dave\superscript{2} \hspace{1mm}
    Connor Henley\superscript{1} \hspace{1mm} \\
    Ashok Veeraraghavan\superscript{2} \hspace{1mm}
    Ramesh Raskar\superscript{1}\\
    \superscript{1}Massachusetts Institute of Technology \hspace{5mm}
    \superscript{2}Rice University\\
    {\tt\small \{sidsoma,co24401,raskar\}@mit.edu, \{akshat,vashok\}@rice.edu}
}

\twocolumn[{%
\renewcommand\twocolumn[1][]{#1}%
\maketitle
\begin{center}
    \centering
    \captionsetup{type=figure}
    \includegraphics[width=0.98\textwidth]{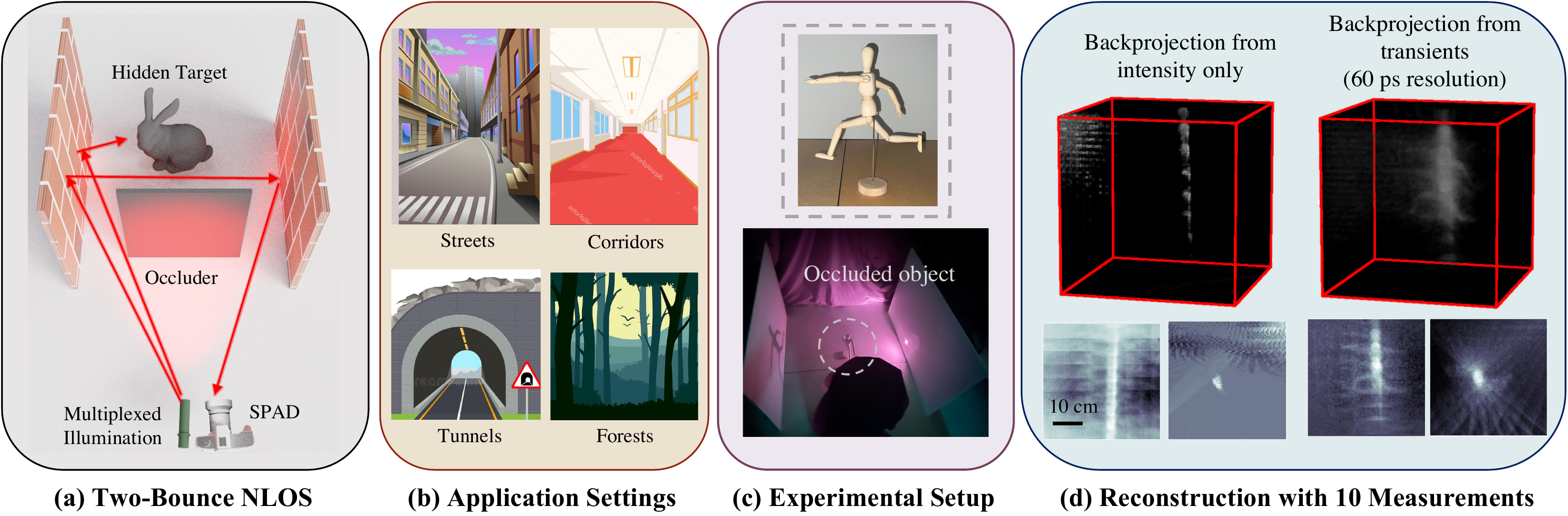}
    \captionof{figure}{\textbf{Two-bounce Transient NLOS Imaging.} (a) We propose using two-bounce transients for NLOS imaging. Two-bounce signals capture information about the shadows in a scene, and use of transient information can reduce the number of measurements needed for reconstruction by using multiplexed illumination. (b) Our work can be applicable in many scenarios, such as tunnels, corridors, streets, and forests. (c)-(d) We experimentally validate the use of two-bounce transients on real objects. We show that time-of-flight information enables more robust reconstructions of occluded objects while using fewer measurements.}
    \label{fig:teaser}
    
\end{center}%
}]
\maketitle

\begin{abstract}
\vspace{-2mm}
\noindent 
The goal of non-line-of-sight (NLOS) imaging is to image objects occluded from the camera's field of view using multiply scattered light. Recent works have demonstrated the feasibility of two-bounce (2B) NLOS imaging by scanning a laser and measuring cast shadows of occluded objects in scenes with two relay surfaces. In this work, we study the role of time-of-flight (ToF) measurements, \ie transients, in 2B-NLOS under multiplexed illumination. Specifically, we study how ToF information can reduce the number of measurements and spatial resolution needed for shape reconstruction. We present our findings with respect to tradeoffs in (1) temporal resolution, (2) spatial resolution, and (3) number of image captures by studying SNR and recoverability as functions of system parameters. This leads to a formal definition of the mathematical constraints for 2B lidar. We believe that our work lays an analytical groundwork for design of future NLOS imaging systems, especially as ToF sensors become increasingly ubiquitous.

\end{abstract}


\section{Introduction}

\label{sec:intro}

\begin{figure}
    \centering
        \includegraphics[width=\linewidth]{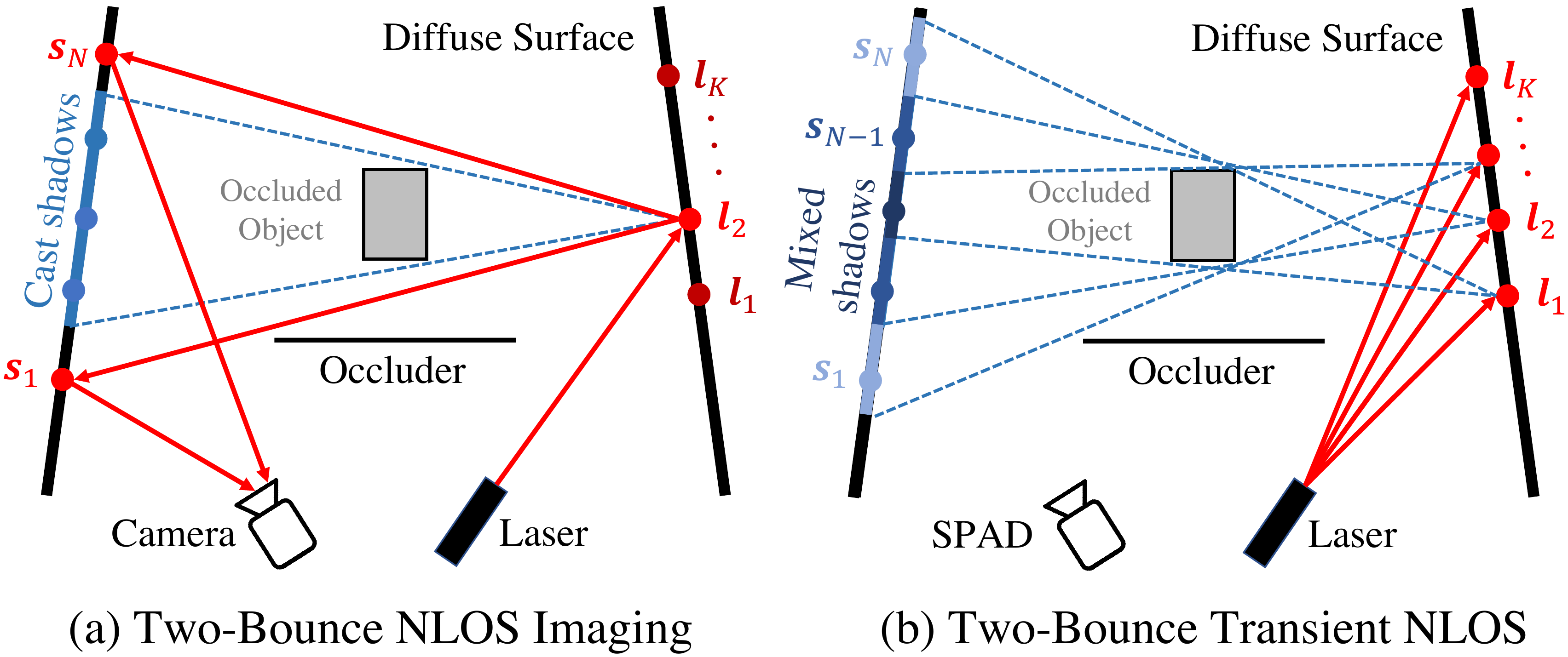}
  \caption{\textbf{Two-Bounce NLOS Imaging.} (a) Two-bounce NLOS imaging performs 3D reconstruction by using information contained in the shadows of the hidden object when illuminated by one virtual source $\mathbf{l}$ at a time. (b) In this work, we study the benefits of using transient information for two-bounce NLOS imaging under the presence of multiplexed illumination.}
  \label{fig:two-bounce-NLOS}
\end{figure}

\begin{figure}
    \centering
\includegraphics[width=0.8\linewidth]{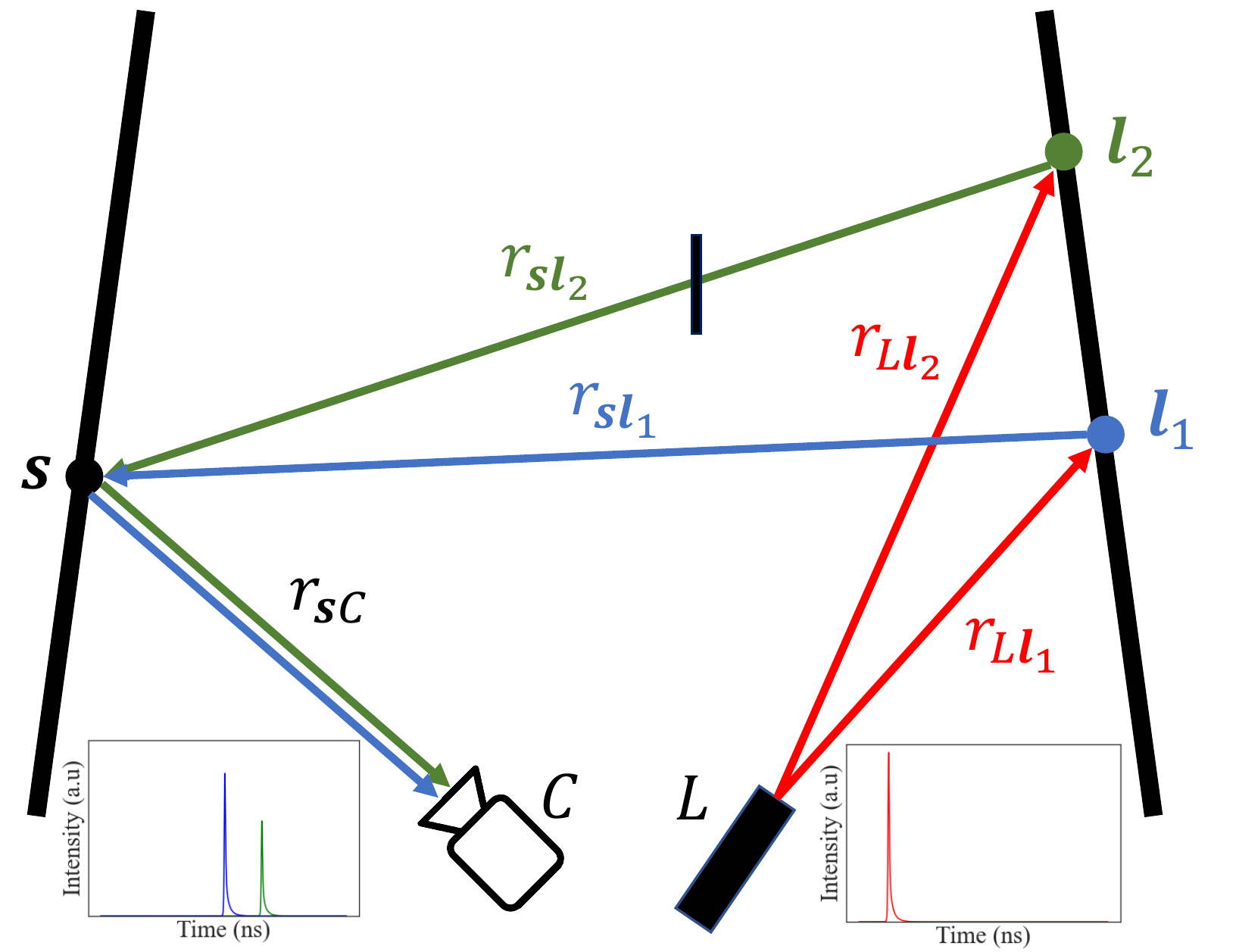}
    \caption{\textbf{Time-of-Flight Information for Demultiplexing Shadows.} 
    Assume the point $\mathbf{s}$ lies in shadow. When using multiplexed illumination, there exists an ambiguity as to whether $\mathbf{s}$ lies in the shadow of $\mathbf{l}_1$, $\mathbf{l}_2$, or both. However, the paths $L\text{-}\mathbf{l}_1\text{-}\mathbf{s}\text{-}C$ and $L\text{-}\mathbf{l}_2\text{-}\mathbf{s}\text{-}C$ have different pathlengths, suggesting the usefulness of ToF information for demultiplexing shadows.}
    \label{fig:tof_benefits}
\end{figure}

Non-line-of-sight (NLOS) imaging aims to reconstruct objects occluded from direct line of sight and has the potential to be transformative in numerous applications across autonomous driving, search and rescue, and non-invasive medical imaging \cite{maeda2019recent}. The key approach is to measure light that has undergone multiple surface scattering events and computationally invert these measurements to estimate hidden geometries. Recent work used two-bounce light, as shown in \cref{fig:two-bounce-NLOS}a, to reconstruct high quality shapes behind occluders \cite{henley2020imaging}. The key idea is that two-bounce  light captures information about the shadows of the occluded object. By scanning the laser source at different points $\mathbf{l}$ on a relay surface and measuring multiple shadow images, it is possible to reconstruct the hidden object by computing the visual hull \cite{laurentini1994visual} of the measured shadows. 
\vspace{-4mm}
\paragraph{Benefits of Two-Bounce}\emph{Two-bounce} (2B) light can be captured using two relay surfaces on opposite sides of the hidden scene (\cref{fig:teaser}a). This can occur in a variety of real-world settings such as tunnels, hallways, streets, and cluttered environments like forests (\cref{fig:teaser}b). For example, imagine an autonomous vehicle in a tunnel being able to see ahead of the car in front of it by using two-bounce signals from the sides of the tunnel. Two-bounce light also helps alleviate the signal-to-noise ratio (SNR) issue in 3B-NLOS imaging, since 3B signals experience significant signal attenuation at each surface scatter event. Furthermore, 2B light captures information about the \emph{cast shadows} of the hidden object instead of reflectance, as in the 3B case. As a result, the SNR of 2B signals is independent of the object's albedo and can therefore robustly image darker objects. 
\vspace{-5mm}
\paragraph{Contributions} In this work, we propose using two-bounce transients for NLOS imaging to enable the use of multiplexed illumination. 3B-NLOS transient imaging systems have been analyzed previously, but we believe we are the first to analyze the use of multiplexed illumination and transient measurements for 2B-NLOS (\cref{fig:two-bounce-NLOS}b). Our key insight is that multiplexed illumination enables measurement of occluded objects with fewer image captures, and 2B transients enable demultiplexing shadows, as shown in \cref{fig:tof_benefits}. The robust SNR and promise of few-shot capture afforded by 2B transients makes the idea an important direction for few-shot NLOS imaging in general environments. We summarize our contributions below.
\begin{itemize}
    \item \textit{Model}: We present new insights into the algebraic formulations of two-bounce lidar for NLOS imaging.
    \item \textit{Analysis}: We analyze tradeoffs with spatial resolution, temporal resolution, and number of image captures by using quantitative metrics such as SNR and recoverability, as well as qualitative validation from real and simulated results.
\end{itemize}
\noindent
Although we envision the ideas introduced here to inspire future works in single-shot NLOS imaging, we do \emph{not} claim this as a contribution in this paper.

\vspace{-4mm}
\paragraph{Scope of This Work} All data used for this work is from simulation and a single-pixel SPAD sensor. We do not physically realize few-shot results in this paper due to limited availability of high-resolution SPAD arrays presently. Instead, we emulate SPAD array measurements by scanning a single-pixel SPAD across the field of view and and emulate multiplexed illumination by scanning a laser spot, acquiring individual transient images, and summing the images in post-processing. These measurements, however, are more conducive to our goal of performing analysis that explores the landscape of possibilities in NLOS imaging with advances in ToF sensors. We believe that the ideas and analysis introduced in this work will be increasingly relevant as SPAD array technology matures \cite{kumagai20217, zhang2021240, morimoto2020megapixel}.

\section{Related Work}

\begin{figure*}
    \centering
    \includegraphics[width=0.9\textwidth]{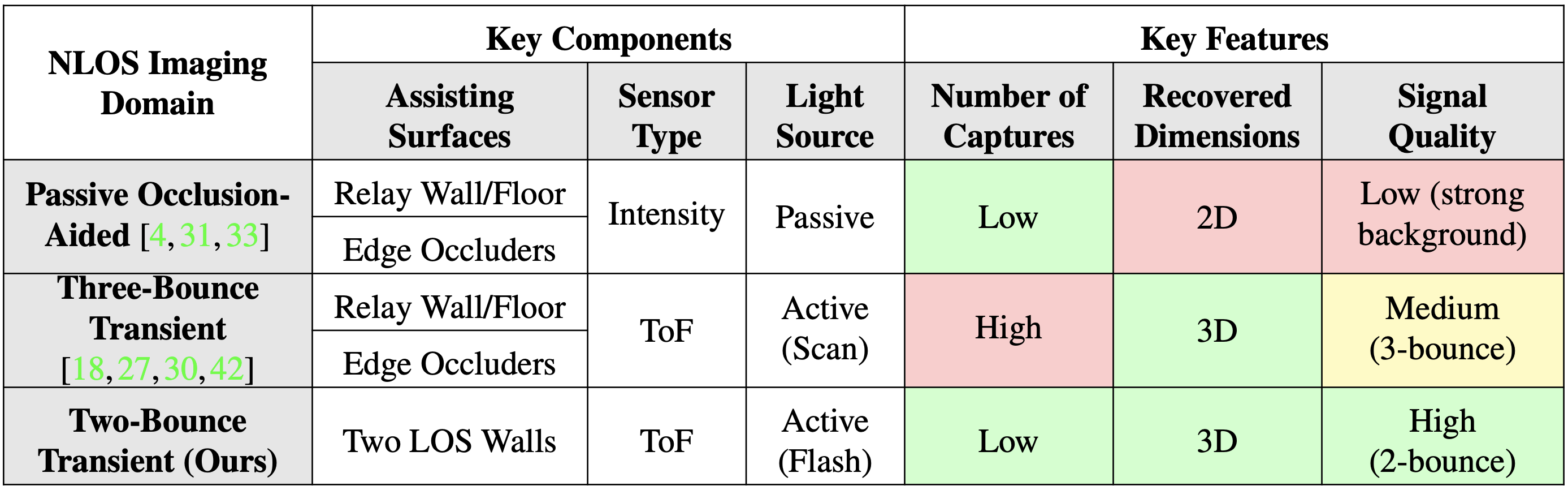}
    \caption{\textbf{Comparison of NLOS Methods.} Passive approaches require fewer measurements but can either recover up to two dimensions or are strongly affected by ambient noise. Works exploiting three-bounce light paths demonstrate 3D imaging around the corner but require a bounce on the hidden surface, reducing the signal quality. We analyze exploiting two-bounce transients for NLOS imaging.}
    \label{table2}
\end{figure*}

\subsection*{Non-Line-of-Sight Imaging}

\paragraph{ToF Methods:}The \textit{non-line-of-sight imaging} (NLOS) problem aims to see around a corner by measuring light reflected off intermediary surfaces. A large body of work models the response of a hidden scene to a temporal pulse of light and generally perform a variant of ellipsoidal tomography or solve a linear inverse problem \cite{kirmani2009looking, velten2012recovering, buttafava2015non, gariepy2016detection, arellano2017fast, heide2019non}. The temporal profiles can probe features of the hidden space, such as Fermat paths \cite{xin2019theory}, scene impulse response \cite{liu2019non}, and individual voxels \cite{pediredla2019snlos}. Other works derive efficient algorithms using a confocal setup \cite{o2018confocal,lindell2019wave, ahn2019convolutional} and data-driven techniques \cite{chen2020learned}. NLOS has also been demonstrated with amplitude modulated continuous wave (AMCW) ToF by exploiting sparsity and fitting measurements to a Lambertian reflectance model \cite{kadambi2016occluded, heide2014diffuse}. 
\vspace{-4mm}
\paragraph{Other Active Methods:}Chen et al. use steady-state (\ie time-integrated) intensity information by leveraging partial directionality of diffuse reflectors \cite{chen2019steady}. Radio wavelengths can also be used to image through walls and/or around them \cite{adib2013see, scheiner2020seeing, yue2022cornerradar}. There has also been significant progress in exploiting speckle correlations \cite{metzler2020deep,dave2020foveated,smith2018tracking,katz2014non} and synthetic wave holography \cite{willomitzer2021fast,willomitzer2019high} for NLOS imaging. Henley et al. demonstrate the use of two-bounce light (\ie shadows) for NLOS using intensity sensors \cite{henley2020imaging}. An extension of this work uses ToF constraints on two-bounce pathlengths to isolate shadows from ambient background and higher-order inter-reflections \cite{henley2022bounce}. However multiplexing was not used.
\vspace{-4mm}
\paragraph{Passive Methods:}Passive methods make use of optical information already present in the scene without an active source. Many passive approaches exploit polarization cues \cite{tanaka2020polarized,dave2022pandora}, ambient thermal light \cite{tanaka2020polarized,lin2020passive}, shadows \cite{swedish2021objects}, multi-view reflections \cite{tiwary2022orca}, and neural networks \cite{sharma2021you, tancik2018flash} to reconstruct or make inferences about the scene outside the camera's field of view. 

\textit{Occlusion-assisted imaging} makes use of information present at sharp discontinuities in LOS and NLOS scenes by turning them into virtual pinholes \cite{bouman2017turning, rapp2020seeing} or virtual pinspecks \cite{saunders2019computational}. Intensity-only methods are difficult to scale to scenarios with arbitrary lighting conditions.

\vspace{-4mm}
\section{Forward and Adjoint Operators}

\subsection{Problem Setup}
We first introduce terminology and notation for our problem formulation. The object of interest lies behind an occluder and is flanked on two sides by planar Lambertian relay surfaces. Without loss of generality, the right wall is the \emph{illumination wall} $\mathcal{L}$ and the left wall is the \emph{observation wall} $\mathcal{O}$. Light incident on the illumination wall at a point $\mathbf{l}$ is a \emph{virtual source}. Each virtual source diffusely radiates light towards the observation wall. Light will arrive at a \emph{virtual detector} $\mathbf{s}$ from $\mathbf{l}$ if the object doesn't lie along the ray connecting $\mathbf{s}$ and $\mathbf{l}$, as shown in \cref{fig:two-bounce-NLOS}a. If the object obstructs this ray, $\mathbf{s}$ is said to lie in the \emph{shadow} of the object. We assume that every virtual source and virtual detector is visible to the laser and camera field of view, respectively. The hidden scene is discretized into a voxel grid described by a binary occupancy (or transparency) function. This implicity assumes that we are reconstructing opaque surfaces. We neglect the effect of higher order bounces (e.g. interreflections, 3B light) because SPADs can temporally gate out the longer ToF of 3B light. Furthermore, the intensity of higher order bounces will be much lower than that of two-bounce light because all surfaces are assumed to be diffuse.

We denote \emph{light transients} as ToF measurements for light that arrives at the virtual detector and \emph{shadow transients} as ToF measurements for light that didn't arrive at the virtual detector because of the object's shadow. Both of these are \emph{two-bounce transients}. We show that modeling the problem in terms of shadow transients enables reconstruction of the hidden objects shape by solving a linear inverse problem.

\subsection{Linear Forward Model}

\begin{figure}
    \centering
\includegraphics[width=0.95\linewidth]{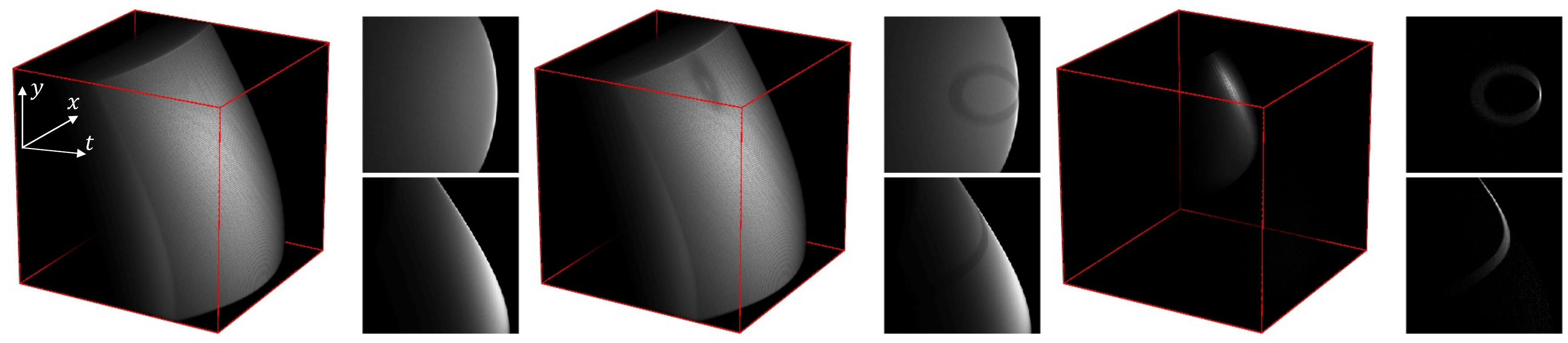}
    \caption{\textbf{Light and Shadow Transients.} Three spacetime curves are depicted. (left) Empty transient $i_0$ with no hidden object, (middle) light transient $i_l$ with an object, and (right) shadow transient $i_s=i_0-i_l$. Switching from light transients $i_l$ to shadow transients $i_s$ enables derivation of a linear forward model.}
    \label{fig:shadow_transients}
\end{figure}

\paragraph{Light Transients} Consider the case where the occluded scene is empty. The measured transient is then given by
\begin{equation}
    i_0(t;\mathbf{s}) = \alpha  \iint_\mathcal{L} \frac{\delta(\|\mathbf{l}-\mathbf{g}\| + \|\mathbf{l}-\mathbf{s}\| - ct)}{\|\mathbf{l}-\mathbf{s}\|^2}d\mathbf{l},
    \label{eq:empty_transient}
\end{equation}
\noindent 
where $\mathbf{g}$ is the location of the laser, $c$ is the speed of light, and $\alpha$ is the intensity of the laser. We refer to $i_0$ as the \emph{empty transient}. For simplicity, we ignore $\alpha$ for the rest of the derivations in this section and assume that the length between the virtual detector and camera is calibrated or known. The numerator of the integrand 
\begin{equation}
    i(t; \mathbf{s}, \mathbf{l}) = \delta(\|\mathbf{l}-\mathbf{g}\| + \|\mathbf{l}-\mathbf{s}\| - ct)
    \label{eq:delta}
\end{equation}
\noindent 
describes the equation of an ellipsoid, with foci at $\mathbf{g}$ and $\mathbf{s}$ and major axis length $ct$. 
When an object is present in the occluded scene, the measured light transient is given by 
\begin{equation}
    i_l(t;\mathbf{s}) = \iint_\mathcal{L} V_{\mathbf{s},\mathbf{l}} \cdot \frac{i(t; \mathbf{s}, \mathbf{l})}{\|\mathbf{l}-\mathbf{s}\|^2}  d\mathbf{l},
\end{equation}
\noindent 
where $V_{\mathbf{s},\mathbf{l}}$ is the visibility between $\mathbf{s}$ and $\mathbf{l}$.
\vspace{-2mm}
\begin{equation}
    V_{\mathbf{s}, \mathbf{l}} = \exp \left( \oint_\mathbf{l}^\mathbf{s}\ln(f_t(\mathbf{x}))d\mathbf{x} \right)
    \label{eq:prod_integral}
\end{equation}
\noindent 
where \cref{eq:prod_integral} computes the product integral along the line connecting $\mathbf{l}$ and $\mathbf{s}$ and $f_t(\mathbf{x}) \in \{0, 1\}$ is the transparency function of voxel $\mathbf{x}$ in the occluded region. 
\vspace{-4mm}
\paragraph{Shadow Transients} The non-linearity of \cref{eq:prod_integral} makes model inversion challenging with light transients. Casting the formulation from light space-time to \emph{shadow space-time} enables a linear formulation of the problem. A shadow transient at $\mathbf{s}$ is given by 
\begin{equation}
    i_s(t; \mathbf{s}) = i_0(t;\mathbf{s}) - i_l(t;\mathbf{s}).
    \label{eq:shadow_transient}
\end{equation}
\noindent 
\cref{fig:shadow_transients} illustrates the difference between light transients and shadow transients. Shadow transients directly probe voxel occupancy, whereas light transients probe for voxel transparency. For example, if the occluded scene is empty except for voxel $\mathbf{x}$, the shadow transient is given by
\begin{equation}
    i_s(t;\mathbf{s}, \mathbf{x}) = \frac{\delta(\|\mathbf{l}(\mathbf{x}, \mathbf{s})-\mathbf{g}\| + \|\mathbf{l}(\mathbf{x}, \mathbf{s})-\mathbf{s}\| - ct)}{\|\mathbf{l}(\mathbf{x}, \mathbf{s})-\mathbf{s}\|^2},
\end{equation}
\noindent 
where $\mathbf{l}(\mathbf{x}, \mathbf{s})$ is a point on the illumination wall that lies along the ray connecting $\mathbf{x}$ and $\mathbf{s}$. If the wall doesn't intersect this ray, $i_s(t; \mathbf{x}, \mathbf{s}) = 0$.

\vspace{-4mm}
\paragraph{Linear Relaxation} Now, consider the general case where a single opaque object lies in the hidden scene. We can describe the shadow transient as
\begin{equation}
    i_s(t; \mathbf{s}) = \iint_\mathcal{L} O_{\mathbf{s}, \mathbf{l}} \cdot \frac{i(t;\mathbf{s}, \mathbf{l})}{\|\mathbf{l}-\mathbf{s}\|^2} d\mathbf{l},
    \label{eq:forward}
\end{equation}
\noindent 
where $O_{\mathbf{s}, \mathbf{l}} \in \{0, 1\}$ is the occlusion function (denotes if the ray between $\mathbf{l}$ and $\mathbf{s}$ is occluded). It can be expressed as the indicator function
\vspace{-2mm}
\begin{equation}
    \begin{split}
        O_{\mathbf{s}, \mathbf{l}} &= 
        \begin{cases}
            1, & \text{if } \oint_\mathbf{l}^\mathbf{s}f_o(\mathbf{\mathbf{x}})d\mathbf{x} > 0 \\
            0,               & \text{otherwise}
        \end{cases} \\
        &\approx \oint_\mathbf{l}^\mathbf{s}f_o(\mathbf{\mathbf{x}})d\mathbf{x},
        \end{split}
    \label{eq:occlusion}
\end{equation}
\noindent 
where $f_o(\mathbf{x})$ is the voxel occupancy function of the hidden scene. The linear relaxation in \cref{eq:occlusion} is possible because NLOS reconstructions typically only consider ToF information, but not radiometric information. The above formulation will preserve the structure of the streak images in space-time, even if the intensities are overestimated. Combining \cref{eq:shadow_transient}, (\ref{eq:forward}), and (\ref{eq:occlusion}), we obtain  
\begin{equation}
    i_0(t;\mathbf{s}) - i_l(t;\mathbf{s}) = \iint_{\mathcal{L}} \frac{i(t; \mathbf{s}, \mathbf{l})}{\|\mathbf{l}-\mathbf{s}\|^2} \cdot \biggl(\oint_\mathbf{l}^\mathbf{s}f_o(\mathbf{\mathbf{x}})d\mathbf{x}\biggr)d\mathbf{l},
    \label{eq:forward_model}
\end{equation}
\noindent 
where we solve for $f_o(\mathbf{x})$. We can estimate $i_0$ from the scene geometry and $i_l$ is the measured transient.
\vspace{-4mm}
\paragraph{Discrete Beam Illumination} Discretizing $\mathcal{L}$ as $K$ discrete laser spots and the hidden space into a voxel grid, we can write \cref{eq:forward_model} as   
\begin{equation}
    i_s[t; \mathbf{s}] = \sum_{k=1}^{K} \frac{\delta[\|\mathbf{l}_k-\mathbf{g}\| + \|\mathbf{l}_k-\mathbf{s}\| - ct]}{\|\mathbf{l}_k-\mathbf{s}\|^2} \cdot \Biggl( \sum_{\mathbf{x} \in \mathcal{P}_{\mathbf{l}_k, \mathbf{s}}} f_o[\mathbf{x}] \Biggr),
    \label{eq:discrete_forward}
\end{equation}
\noindent 
where $\mathcal{P}_{\mathbf{l}, \mathbf{s}}$ denotes the set of voxels lying along the path between $\mathbf{l}$ and $\mathbf{s}$. \cref{eq:discrete_forward} can be expressed in matrix form  
\begin{equation}
    \mathbf{I}_s= \mathbf{P} \mathbf{B} \mathbf{f} = \mathbf{A}\mathbf{f},
    \label{eq:forward_model_discrete}
\end{equation}
\noindent 
where $\mathbf{I}_s \in \mathbb{R}^{n_un_vn_t}$ is the vectorized space-time measurements, $\mathbf{f} \in \mathbb{R}^{n_xn_yn_z}$ is the voxel occupancies, and $\mathbf{A} \in \mathbb{R}^{n_un_vn_t \times n_xn_yn_z}$ is the measurement operator mapping voxel occupancies to space-time measurements. $n_u$ and $n_v$ are the number of pixels along the $u$ and $v$ direction, and $n_t$ is the number of timing bins. $n_x$, $n_y$, and $n_z$ are the number of voxels in the $x$, $y$, and $z$ directions. 

$\mathbf{P}$ and $\mathbf{B}$ model the temporal and spatial relationships, respectively, between virtual sources, virtual detectors, and voxels. $\mathbf{P} \in \mathbb{R}^{n_un_vn_t \times n_un_vK}$ is a block diagonal matrix, with the column space of the $i$th block $\mathbf{P}^i \in \mathbb{R}^{n_t \times K}$ containing the corresponding $\delta(\cdot)$ functions from \cref{eq:delta} for virtual detector $\mathbf{s}_i$ and all $K$ virtual sources. $\mathbf{B} \in \mathbb{R}^{n_un_vK \times n_xn_yn_z}$ is  composed of $n_un_v$ vertical sub-blocks. Each entry $\mathbf{B}_{kj}^i \in \{0, 1\}$ of the $i$th sub-block $\mathbf{B}^i \in \mathbb{R}^{K \times n_xn_yn_z}$ of sub-block $i$ indicates whether voxel $j$ lies along the line connecting virtual detector $\mathbf{s}_i$ and virtual source $\mathbf{l}_k$. The multiplication $\mathbf{B}\mathbf{f}$ performs the line integral in \cref{eq:occlusion}. The $k$th column vector of $\mathbf{P}$ is the space-time impulse response (STIR) to the $k$th illumination source when the hidden scene is completely occupied (note that we are dealing with shadow transients, not light transients). The $i$th column of $\mathbf{A}$ forms the expected space-time measurements if only the $i$th voxel is occupied (\ie the STIR to a single voxel). We neglect the $r^2$ falloff term in our reconstructions. Additional explanation of the $\mathbf{A}$ matrix is provided in the supplementary. The linear model in \cref{eq:forward_model_discrete} resembles that of 3B-bounce NLOS, where $\mathbf{A}$ is a measurement operator that maps voxels $\mathbf{f}$ to measurements $\mathbf{I}$. However, in 3B-NLOS, the information is encoded into the \emph{reflectance} of the object, whereas in 2B-NLOS, the information is encoded into the \emph{shadows}. Therefore, the former is linear with respects to light transients, and the latter is linear with respect to shadow transients. 

\subsection{Reconstruction Algorithm}

Reconstruction of $\mathbf{f}$ can be performed by backprojection
\vspace{-2mm}
\begin{equation}
    \mathbf{f}_{bp} = \mathbf{A}^\top \mathbf{I}_s.
    \label{eq:backprojected}
\end{equation}

\noindent 
Backprojection produces a low-frequency reconstruction of $\mathbf{f}$. In practice, filtering $\mathbf{f}_{bp}$ with a Laplacian filter 
\begin{equation}
    \mathbf{f}_{fbp} = \Bigg( \frac{\partial^2}{\partial x^2} + \frac{\partial^2}{\partial y^2} + \frac{\partial^2}{\partial z^2} \Bigg) \mathbf{f}_{bp}
\end{equation}
\noindent  
has been shown to yield reasonably good results in ellipsoidal tomography in 3-bounce NLOS and line-based tomography in medical imaging. Our linear formulation also enables reconstruction with priors. We recognize that these optimization-based methods would improve reconstruction over unfiltered backprojection, which is equivalent to a low-frequency version of voxel carving in our setup. However, we restrict our focus to unfiltered backprojection because of its convenient relationship with the Gram matrix, which enables further analysis of our method in \cref{sec:recoverability}. 

\section{Design and Analysis}
In this section, we analyze how the performance of the proposed two-bounce transient system varies with system parameters and the geometry of LOS relay surfaces. We also analyze the impact of temporal resolution, spatial resolution, and number of measurements on recoverability. The amount of multiplexing in the scene is inversely related to the number of measurements. For example, if $60$ laser spots are used and $15$ measurements are captured, then each image contains $4$ multiplexed shadows. We use SNR, coherence, and FWHM of the PSF to quantify performance. 

\begin{figure}
    \centering
    \includegraphics[width=0.45\textwidth]{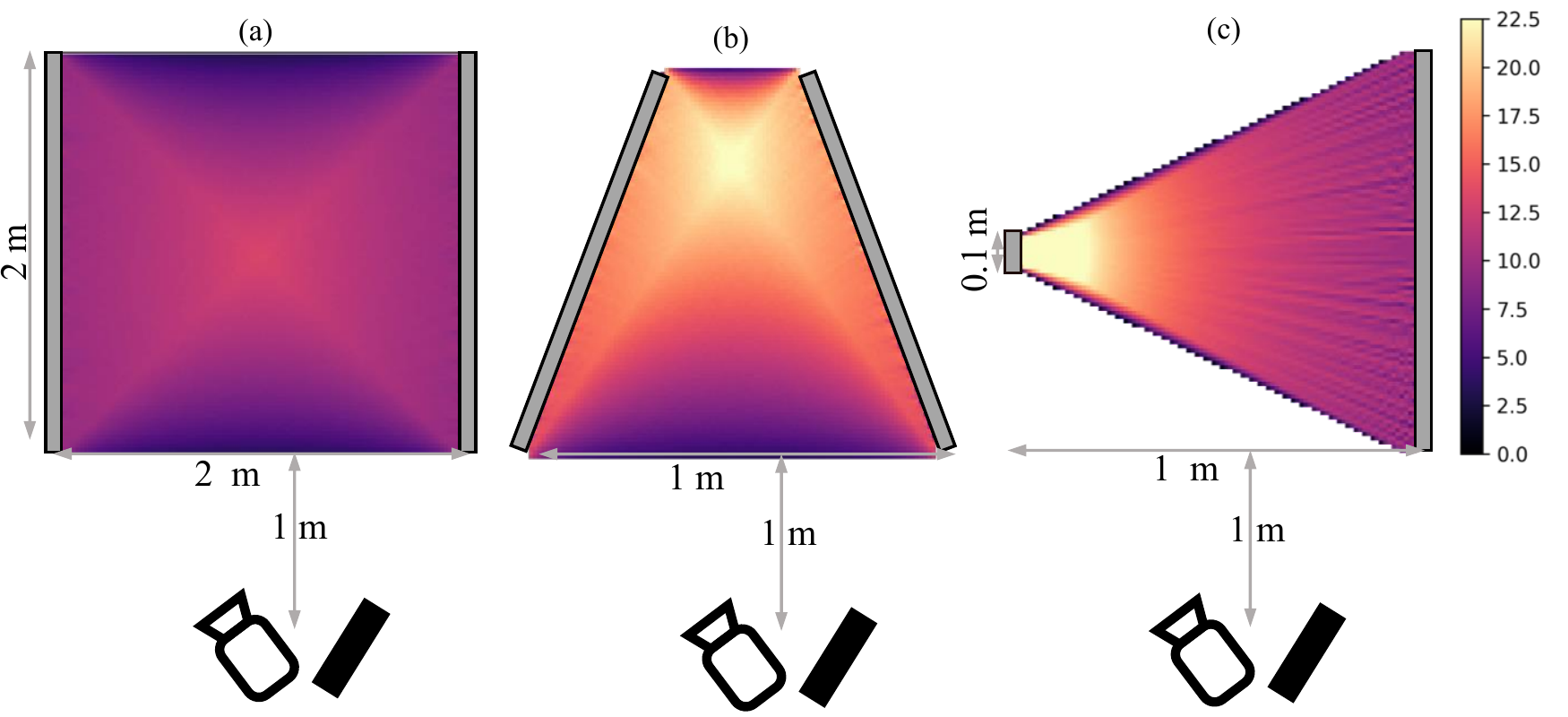}
    \caption{\textbf{SNR Analysis.} The goal of this analysis is to determine how scene geometry affects the SNR of the two-bounce measurement at each voxel in the hidden scene. The first two cases consider the orientations of the relay surfaces, and the third considers the field of view of the camera. SNR is highest in the middle of the hidden scene for parallel surfaces, orienting the walls closer together increases SNR due to reduced $r^2$ falloff, and SNR is highest closer to the virtual detectors for small FOV cameras. }
    \label{fig:snr_analysis}
\end{figure}

\subsection{Voxel SNR Analysis}
We quantify the noise characteristics of the measured shadow two-bounce transients using per-voxel SNR. Consider the case where only voxel $\mathbf{x}$ is occupied in the hidden region. We derive the SNR of the number of ''shadow'' photons received in the measured shadow transient $\mathbf{I}_s^\mathbf{x}$. 
In vector form, the occupancy function can be described as $\mathbf{\boldsymbol\delta}^{\mathbf{x}}$. From \cref{eq:forward_model_discrete}, the shadow transient is given by
\begin{equation}
   \mathbf{I}_s^{\mathbf{x}}  = \mathbf{A}\mathbf{\boldsymbol\delta}^{\mathbf{x}}
\end{equation}
and the light transients measured by the SPAD sensor are given by $\mathbf{I}_l^\mathbf{x}$. For this analysis, we account for $r^2$ falloff because it is a physically accurate phenomenon that impacts SNR. From \cref{eq:shadow_transient}, 
\begin{equation}
    \mathbf{I}_l^\mathbf{x} = 
    \mathbf{I}_0 - \mathbf{A}\mathbf{\boldsymbol\delta}^{\mathbf{x}}
\end{equation}
where $\mathbf{I}_0$ is the empty transient and assumed to be calibrated. Assuming no SPAD pile-up distortion, the number of photons arriving at the sensor for each time bin of $\mathbf{I}_l^\mathbf{x}$ follow an independent Poisson distribution \cite{pediredla2018signal}. Thus, the total number of photons $P^\mathbf{x}$ measured follows a Poisson distribution with the rate parameter being the sum of all the photon counts for all time bins for all pixels.
\begin{equation}
    P^{\mathbf{x}} \sim \text{Poisson}(\mathbf{1}^T(\mathbf{I}_0 - \mathbf{A}\mathbf{\boldsymbol\delta}^{\mathbf{x}}))
\end{equation}
 where $\mathbf{1} \in \mathbb{R}^{n_un_vn_t}$ is the one vector. Using the fact that SNR of a Poisson distribution is $\mu/\sigma = \sqrt{\lambda}$ with rate parameter $\lambda$, the per-voxel SNR $S^{\mathbf{x}}$ is given by  
 \begin{equation}
    S^{\mathbf{x}} =  \sqrt{\mathbf{1}^T(\mathbf{I}_0 - \mathbf{A}\mathbf{\boldsymbol\delta}^{\mathbf{x}})}
\end{equation}
The per-voxel SNR is plotted in \cref{fig:snr_analysis} for three different scenarios. The first two scenarios deal with the orientation of the walls, and the third considers the field of view of the camera. In \cref{fig:snr_analysis}a, we can see that the SNR is higher for center voxels, as more rays connecting the illumination wall and observation wall pass through the center voxels. When the relay walls are oriented towards each other (\cref{fig:snr_analysis}b), the SNR increases due to reduced $r^2$ falloff. When the camera's field of view is small compared to the laser (\cref{fig:snr_analysis}c), regions closer to the observation wall have the highest SNR.

\begin{figure}
    \centering
    \includegraphics[width=0.42\textwidth]{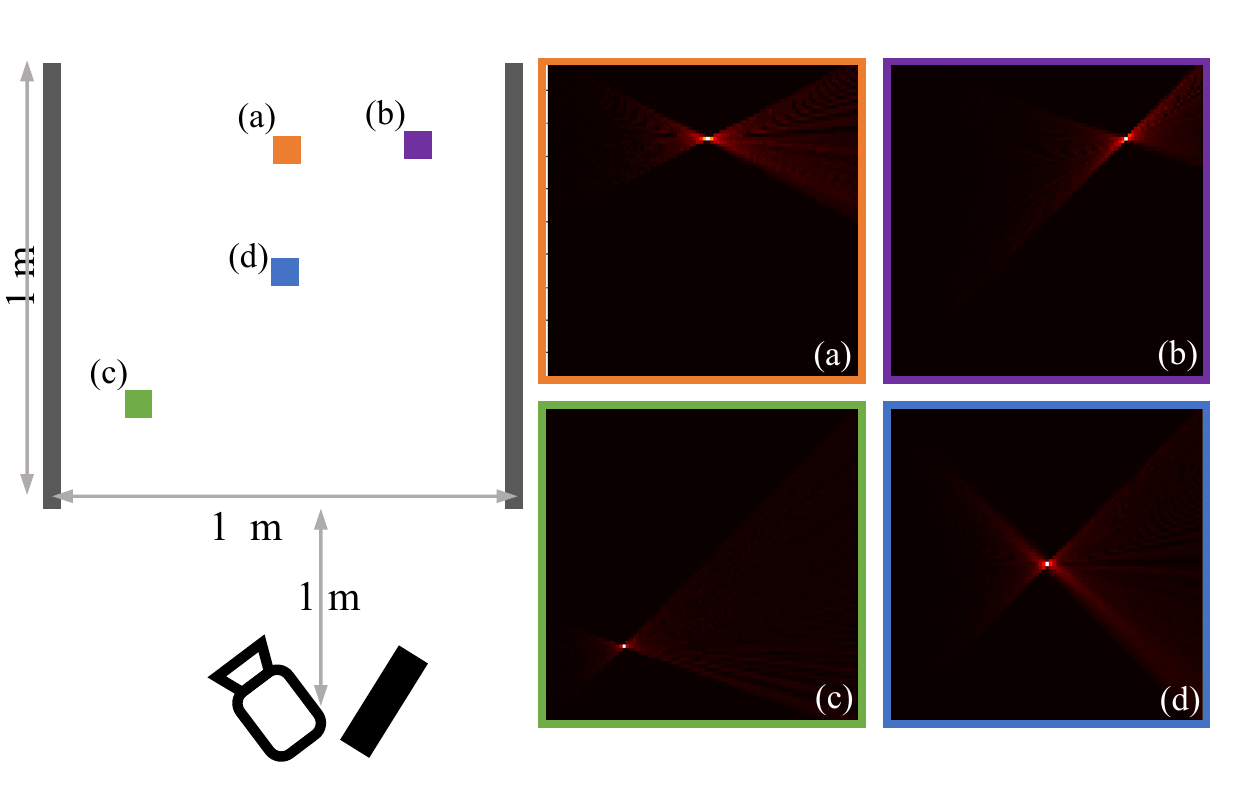}
    \caption{\textbf{Spatially Varying PSF.} In two-bounce lidar, the PSF has a double cone shape. The orientation of the double cone varies strongly as a function of the orientation and length of the relay surfaces (\ie the virtual apertures) and the voxel position.}
    \label{fig:resolution_analysis}
\end{figure}

\begin{figure*}[t]
    \centering
    \includegraphics[width=\textwidth]{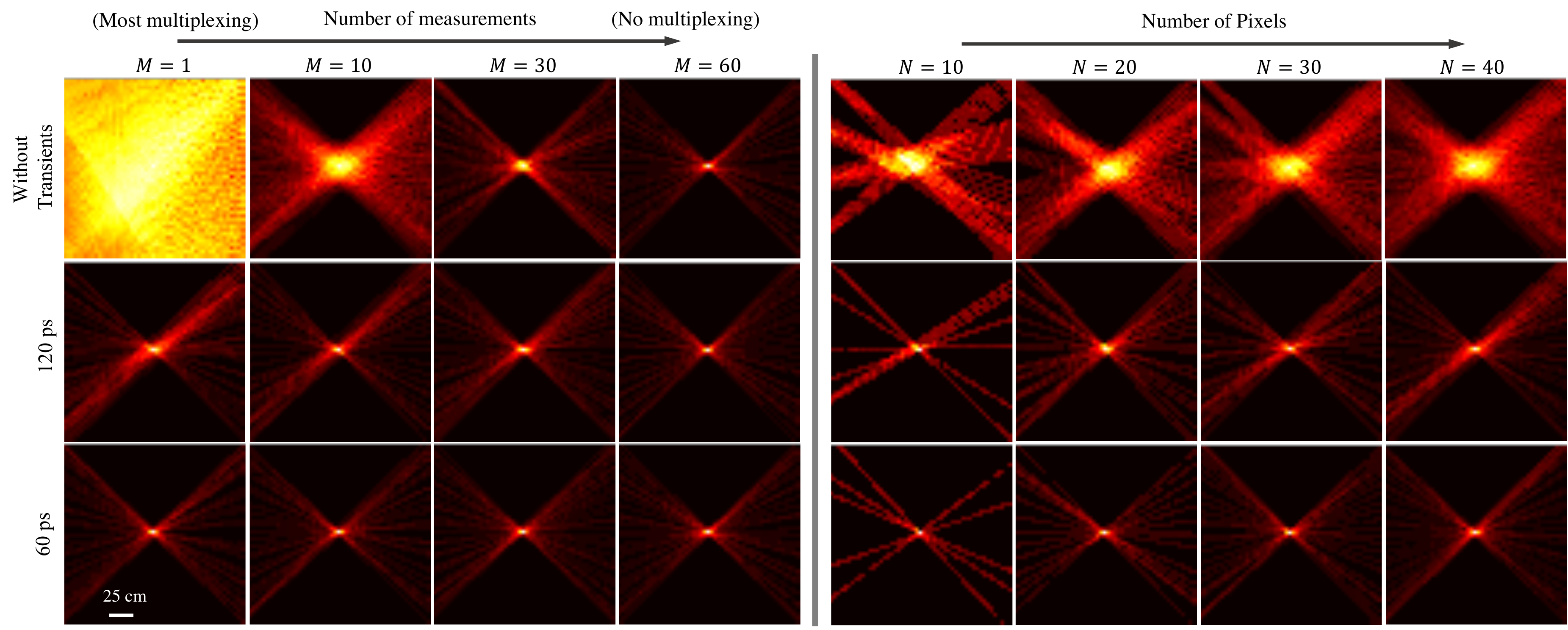}
        \captionof{figure}{\textbf{PSF as a Function of Number of Measurements and Spatial Resolution.} The PSF of the center voxel is shown in the transient and intensity-only case when multiplexed with $60$ laser spots. (1) the number of pixels $N=40$ are kept fixed with the number of measurements $M$ being swept (left) and (2) $M=10$ is kept fixed and $N$ is swept (right). Transient measurements result in PSFs with narrower FWHMs, even with low spatial resolution and low number of measurements.} 
        \label{fig:PSF_analysis}
\end{figure*}

\subsection{Recoverability Analysis}
\label{sec:recoverability}

\paragraph{Resolution}
We now study the resolution limits of our method. In particular, we consider two key metrics in our measurement operator: mutual coherence and full width at half maximum (FWHM) of the PSF. We are interested in how (1) temporal resolution, (2) spatial resolution $N$, and (3) number of measurements $M$ impact each of our metrics. From \cref{eq:backprojected}, we know that the backprojected occupancy can be obtained from the occupancy function by
\begin{equation}
    \mathbf{f}_{bp} = (\mathbf{A}^\top\mathbf{A})\mathbf{f},
\end{equation}
\noindent 
where $\mathbf{G}=\mathbf{A}^\top\mathbf{A}$ is the Gram matrix. We can interpret the columns (or rows) of the Gram matrix to be the spatially varying point spread function (PSF) of the imaging system, as shown in \cref{fig:resolution_analysis}. Each column $\mathbf{G}_i$ describes how a voxel $\mathbf{x}_i$ is blurred during backprojection reconstruction. 

\begin{figure}[t]
    \centering
    \includegraphics[width=0.45\textwidth]{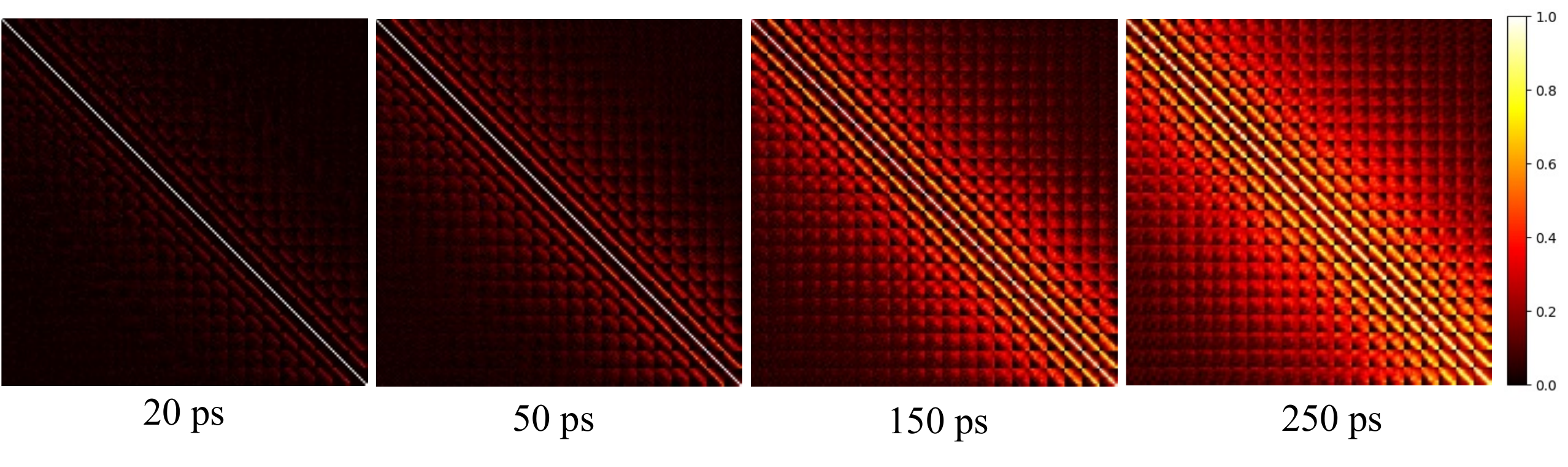}
    \caption{\textbf{Gram Matrix as a Function of Temporal Resolution.} For a defined voxel resolution and fixed spatial resolution. The columns (or rows) of the Gram matrix form the spatially varying PSFs shown in \cref{fig:resolution_analysis}. The off-diagonal streaks describe the coherence between different voxels. As expected, higher temporal resolution decreases the coherence of the Gram matrix, which increases the likelihood of guaranteed sparse recovery.}
    \label{fig:gram_analysis}
\end{figure}

We can see the effect of different imaging parameters on the PSF in \cref{fig:PSF_analysis}, where we sweep different values of (1) $M$ and (2) $N$. In particular, we note that the lobes of the PSF are suppressed when using temporal information, even with low spatial resolution and few number of image measurements. When using no temporal information, the PSF is still roughly centered at the correct location, but has large FWHM. The sharpness of the PSF when using temporal information indicates that it is better suited to applications that require finer voxel resolution with few-shot capture. In case (1) without transients, we see that the FWHM of the PSF is large along both the $x$ and $y$ direction for  small values of $M$. As $M$ approaches the number of illumination spots, the FWHM approaches the theoretical lower bound. In case (2), increased spatial resolution helps better localize the center of the PSF, but the FWHM does not improve. The transient cases are quite robust to both low number of measurements and low spatial resolution. Intuitively, higher values of $N$ in the transient case slightly help accentuate the center voxel by projecting more diverse rays through the center voxel.
\vspace{-4mm}
\paragraph{Mutual Coherence} We wish to further quantify the impact of spatio-temporal resolution and number of measurements on the recoverability of the hidden scene. We know from compressive sensing theory that sparse recovery of the hidden scene is possible when the mutual coherence of $\mathbf{A}$ is sufficiently small. The mutual coherence is defined as 
\begin{equation}
    \mu = \max_{1\leq i \neq j \leq n} \frac{\vert \mathbf{A}_i^\top \mathbf{A}_j\vert}{\vert\mathbf{A}_i\vert \vert \mathbf{A}_j\vert}.
\end{equation}
\noindent 
Intuitively, the coherence measures the largest "similarity" or correlation between any two column vectors in measurement matrix $\mathbf{A}$ \cite{candes2008introduction}. Physically, this quantifies how similar the STIRs of two different voxels are. If $\mu=1$, then two voxels have the exact same STIR, which means that there exists an ambiguity as to which voxel resulted in a certain spatio-temporal response. Alternatively, $\mu$ can be defined as the maximum off-diagonal entry of the Gram matrix. \cref{fig:gram_analysis} illustrates how the Gram matrix (and subsequently the mutual coherence) is affected by coarser timing resolutions. From compressive sensing theory, we know that a $K$-sparse scene can be recovered if $K \leq 0.5(1 + 1/\mu$), which motivates our empirical analysis of the effect of spatial resolution and temporal resolution on $\mu$. 

\begin{figure}
    \centering
    \includegraphics[width=0.32\textwidth]{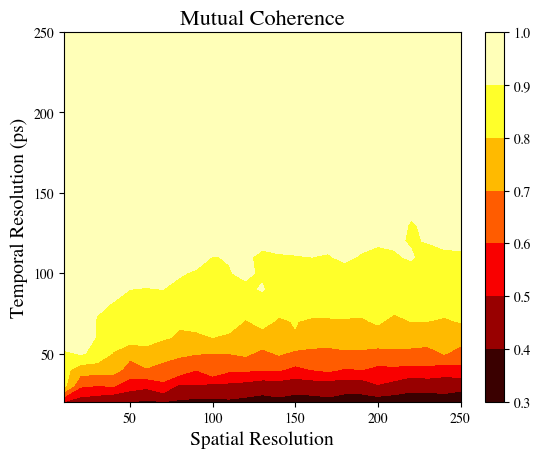}
    \caption{\textbf{Tradeoffs Between Spatial Resolution and Temporal Resolution.} We quantify the tradeoff using mutual coherence. We use the same setup as \cref{fig:resolution_analysis}, but the walls have length $3$ m, the walls are space $2$ m apart, and the distance to the walls is $2$ m. We discretize the voxel resolution to $2$ cm ($\sim 67$ ps). The bottom right of the figure, where spatial and temporal resolution are highest is where mutual coherence is the lowest. For a fixed temporal resolution, improvements in coherence quickly plateau. Improvements in temporal resolution dominate improvements in spatial resolution when reducing coherence, suggesting the importance of ToF for demultiplexing illumination in two-bounce NLOS imaging.}
    \label{fig:coherence_ST}
\end{figure}

From \cref{fig:coherence_ST}, we can see the impact that spatial and temporal resolution have on mutual coherence in a single-shot setup with multiplexed illumination. Consider a scene similar to that of \cref{fig:resolution_analysis}, where the walls are $2$ m apart, the walls are $3$ m long, the axial distance to the walls is $2$ m, and the voxel resolution is discretized to $2$ cm. Increasing the spatial resolution at a fixed temporal resolution appears to have little to no effect on the mutual coherence. As Kadambi et al. also point out, improvements of coherence through higher spatial resolution are limited by the size of the virtual aperture (\ie the wall) \cite{kadambi2016occluded}. On the other hand, the difference between $100$ ps resolution and $10$ ps resolution reduces the coherence from $~0.9$ to $~0.3$ for a fixed spatial resolution. Any timing resolution coarser than $\sim120$ ps results in a coherence of $1$ for multiplexed measurements because pathlength differences for different shadows are no longer resolvable. This analysis motivates the importance of temporal resolution when using multiplexed illumination.



\section{Experiments and Results}

\begin{figure}[t]
    \centering
    \includegraphics[width=0.45\textwidth]{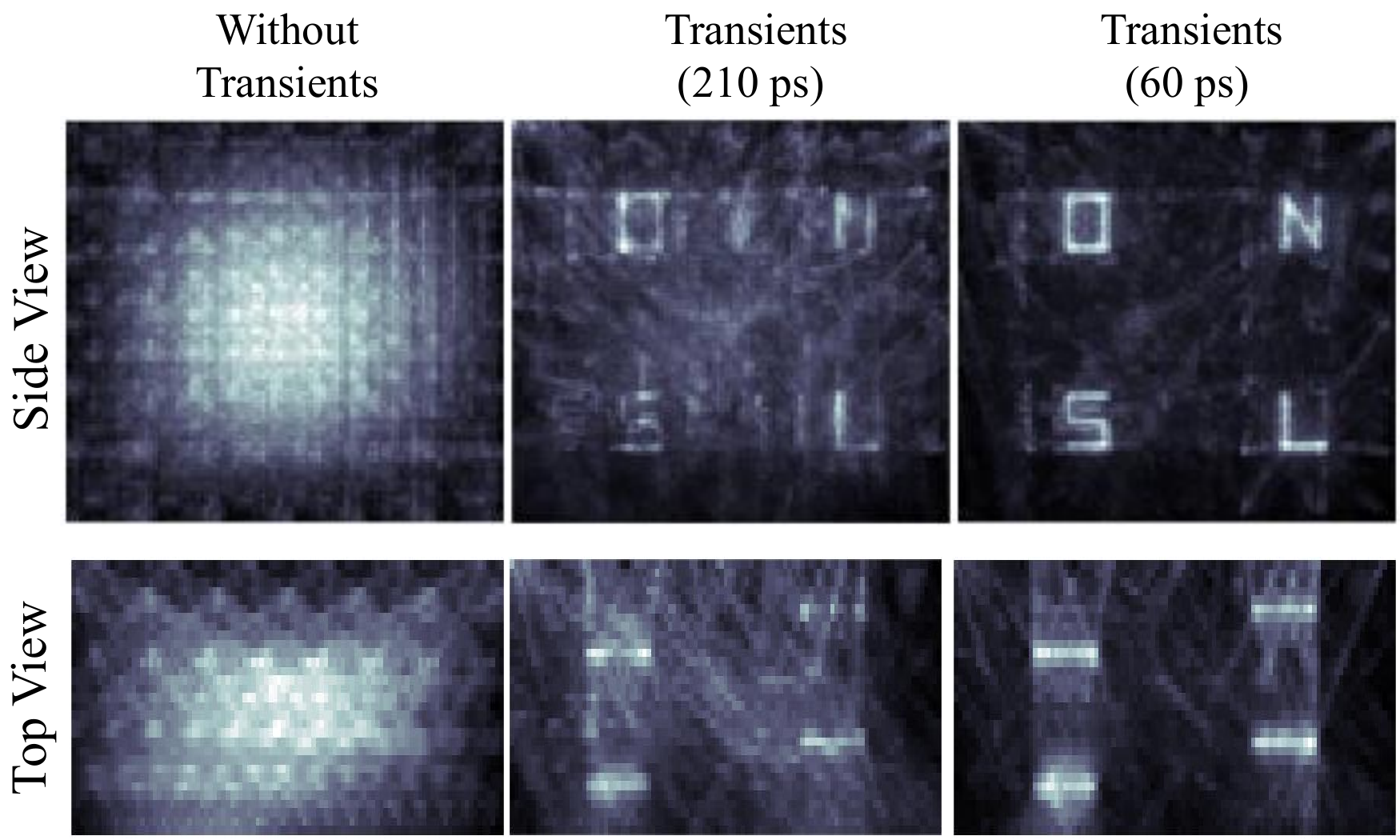}
    \captionof{figure}{\textbf{Single-Shot Localization.} We obtain single-shot reconstructions of the "N", "L", "O", and "S" letters in the hidden scene with and without transient information. We see that using transient information, we are able to effectively localize the hidden object and even get a sharp outline of the letters, demonstrating the promise of single-shot capture with two-bounce transients.} 
    \label{fig:simulated_letters}
\end{figure}
\begin{figure}[t]
    \centering
    \includegraphics[width=0.45\textwidth]{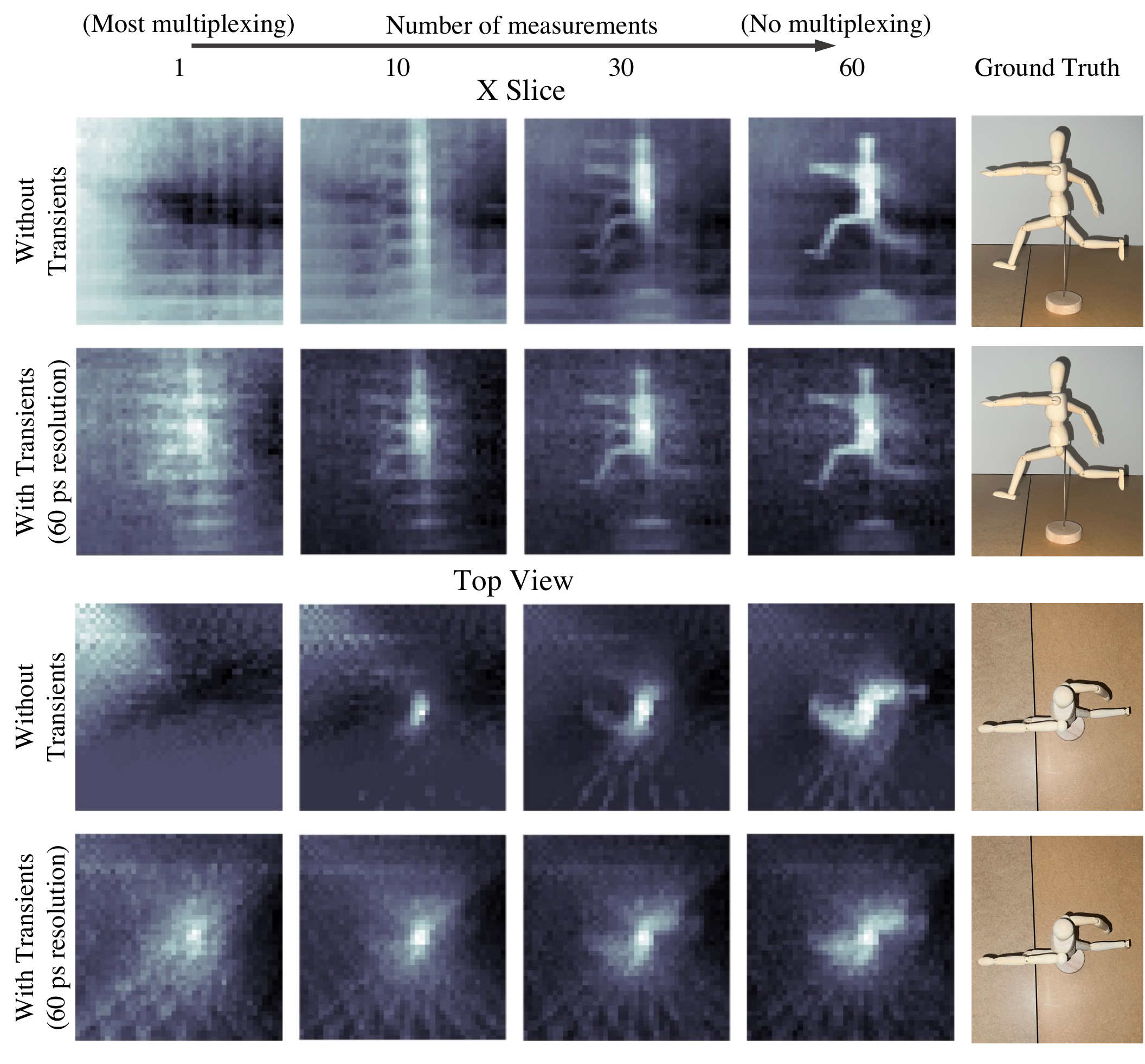}
    \caption{\textbf{Transients Enable Scene Recovery from Multiplexed Illumination.} We perform backprojection of measurements with varying levels of multiplexing. Plotted in the top two rows are $x$ slices of the reconstructed volume. Under no multiplexing (right), intensity is sufficient to carve the shadow along detected pixel. But reconstruction quality sharply deteriorates as we move towards the multiplexing regime (left). When measuring with transients, time-of-flight information allows us to disambiguate between different shadows and recover a rough silhouette of the object even with just a single measurement.}
    \label{fig:real_analysis}
\end{figure}
\begin{figure}[t]
    \centering
    \includegraphics[width=0.45\textwidth]{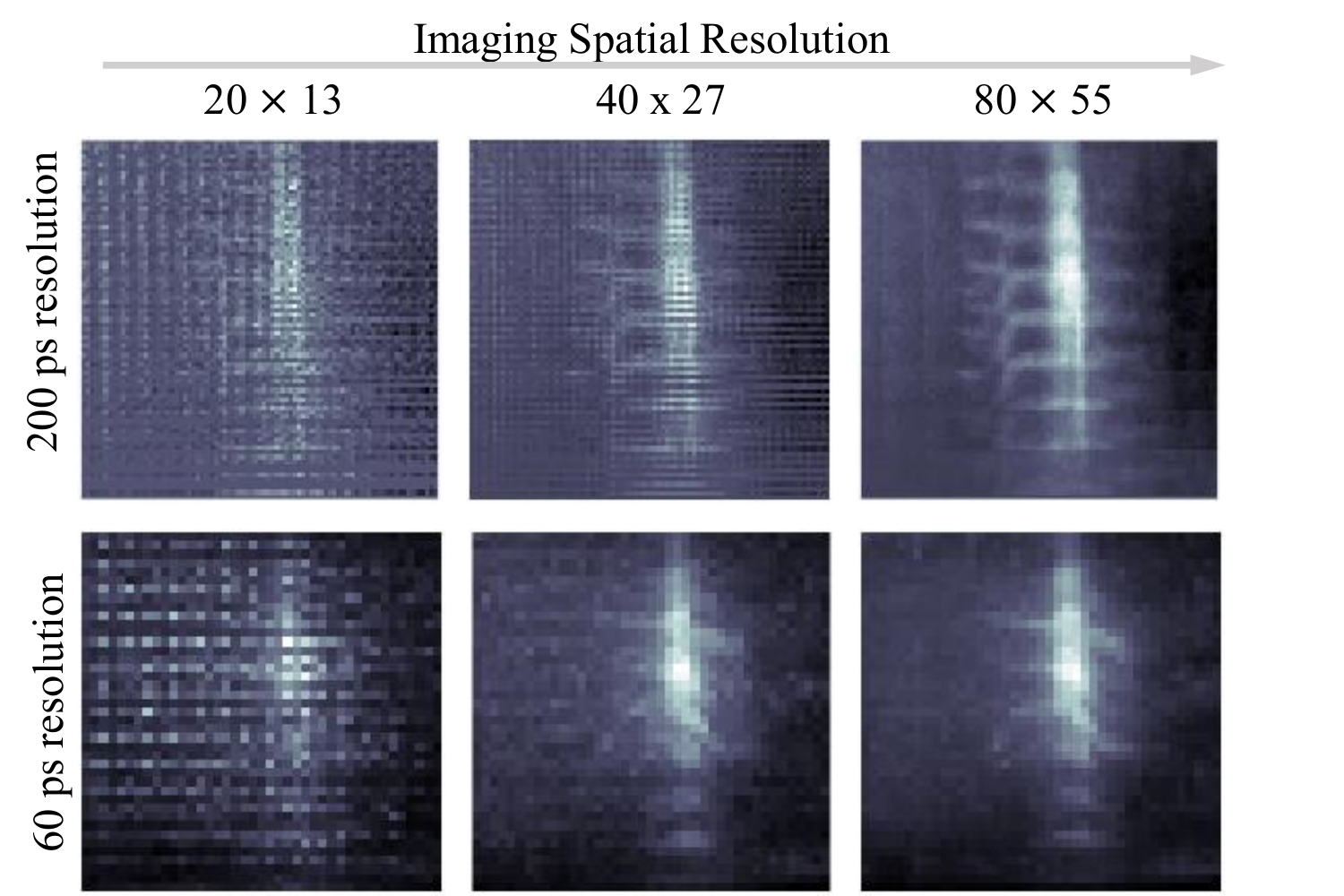}
\caption{\textbf{Spatial Resolution vs. Timing Resolution.} We plot reconstructions with different spatio-temporal resolutions. Lower timing resolution results in aliasing artifacts (\ie repeated arms and legs). Lower spatial resolution introduces hole artifacts.}
    \label{fig:real_analysis_tradeoffs}
\end{figure}

\begin{figure}[t]
    \centering
    \includegraphics[width=0.45\textwidth]{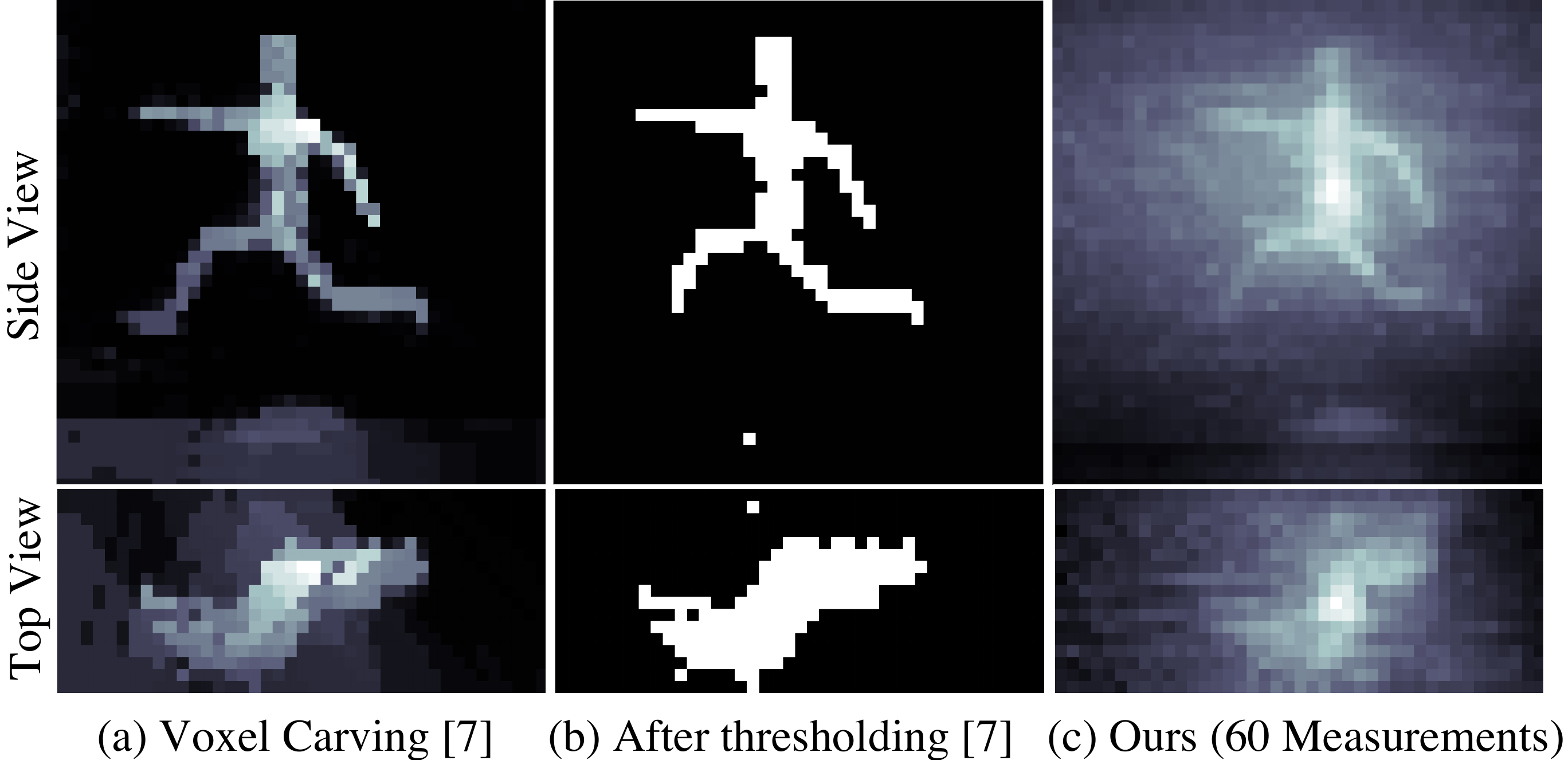}
    \caption{\textbf{Scanning Baseline Comparison.} We compare to the voxel carving method used in \cite{henley2020imaging} (a) before and (b) after thresholding when scanning $1$ laser spot. Our method is more general, but voxel carving yields sharper results with laser scanning.}
    \label{fig:voxel_carving_baseline}
\end{figure}
We scan a single-pixel SPAD across the field of view on the left wall to mimic the output of a SPAD array. We scan a single laser spot across the right wall, and sum up the corresponding transient images in post-processing to mimic the effect of multiplexing. Further details are provided in the supplementary. We first validate our model using simulated data in \cref{fig:simulated_letters}, where we demonstrate that we can localize four letters and reconstruct their 2D shape with just a single capture. We render $i_o$ and $i_l$ to obtain a noiseless version of the shadow transient for this experiment. Then, we reconstruct the shapes of real mannequins and must estimate $i_0$. \cref{fig:real_analysis} shows that ToF measurements provide a noticeable improvement in image reconstruction. When there's no multiplexing (\ie the number of laser spots is the same as number of measurements), the intensity-based method is similar to the ToF-based method. However, in the intensity-only case, the object becomes increasingly blurry until it is no longer visible as the amount of multiplexing increases. In the ToF case for single-shot, we can see a blurry image roughly localized where the center of the mannequin is as well as some high-frequency detail. Furthermore, the top view of the ToF reconstructions show that the high-frequency shape of the arms can be reconstructed even with multiplexed illumination, indicating the promise of single-shot capture with two-bounce ToF information. \cref{fig:real_analysis_tradeoffs} shows that poorer temporal resolution results in aliasing artifacts (repeated arms and legs), whereas the rough shape of the object can still be seen with lower spatial resolution. Finally, we compare our method to the probabilistic voxel carving method used in \cite{henley2020imaging} in \cref{fig:voxel_carving_baseline}. Voxel carving results in sharper reconstructions than backprojection. However, our method can be improved by filtering or incorporating priors, and is also generalizable to multiplexed illumination by utilizing ToF information.

\section{Discussion}
In this paper, we introduced the use of two-bounce transients for NLOS imaging. Compared to 3B-NLOS, 2B-NLOS enables more robust imaging of occluded objects in the presence of scenes with arbitrary lighting conditions and low albedo objects. Transient 2B-NLOS is promising for few-shot capture of occluded objects by using multiplexed illumination, especially as SPAD array technology matures. When comparing tradeoffs in temporal resolution, spatial resolution, and number of measurements, we qualitatively and quantitatively verify with real and simulated results that temporal resolution is most important for demultiplexing shadows. 
We anticipate future work to extend the analysis in this paper to SPAD arrays for real-time NLOS imaging.

\section*{Acknowledgements}
AD and AV are supported by NSF Expeditions Award \#IIS-1730574 and NSF CAREER Award \#IIS-1652633. CH was supported by a Draper Scholarship.

{\small
\bibliographystyle{ieee_fullname}
\bibliography{egbib}

\begin{thebibliography}{10}\itemsep=-1pt

\bibitem{adib2013see}
Fadel Adib and Dina Katabi.
\newblock See through walls with wifi!
\newblock In {\em Proceedings of the ACM SIGCOMM 2013 conference on SIGCOMM},
  pages 75--86, 2013.

\bibitem{ahn2019convolutional}
Byeongjoo Ahn, Akshat Dave, Ashok Veeraraghavan, Ioannis Gkioulekas, and
  Aswin~C Sankaranarayanan.
\newblock Convolutional approximations to the general non-line-of-sight imaging
  operator.
\newblock In {\em Proceedings of the IEEE/CVF International Conference on
  Computer Vision}, pages 7889--7899, 2019.

\bibitem{arellano2017fast}
Victor Arellano, Diego Gutierrez, and Adrian Jarabo.
\newblock Fast back-projection for non-line of sight reconstruction.
\newblock In {\em ACM SIGGRAPH 2017 Posters}, pages 1--2. Optica Publishing
  Group, 2017.

\bibitem{bouman2017turning}
Katherine~L Bouman, Vickie Ye, Adam~B Yedidia, Fr{\'e}do Durand, Gregory~W
  Wornell, Antonio Torralba, and William~T Freeman.
\newblock Turning corners into cameras: Principles and methods.
\newblock In {\em Proceedings of the IEEE International Conference on Computer
  Vision}, pages 2270--2278, 2017.

\bibitem{buttafava2015non}
Mauro Buttafava, Jessica Zeman, Alberto Tosi, Kevin Eliceiri, and Andreas
  Velten.
\newblock Non-line-of-sight imaging using a time-gated single photon avalanche
  diode.
\newblock {\em Optics express}, 23(16):20997--21011, 2015.

\bibitem{candes2008introduction}
Emmanuel~J Cand{\`e}s and Michael~B Wakin.
\newblock An introduction to compressive sampling.
\newblock {\em IEEE signal processing magazine}, 25(2):21--30, 2008.

\bibitem{chen2019steady}
Wenzheng Chen, Simon Daneau, Fahim Mannan, and Felix Heide.
\newblock Steady-state non-line-of-sight imaging.
\newblock In {\em Proceedings of the IEEE/CVF Conference on Computer Vision and
  Pattern Recognition}, pages 6790--6799, 2019.

\bibitem{chen2020learned}
Wenzheng Chen, Fangyin Wei, Kiriakos~N Kutulakos, Szymon Rusinkiewicz, and
  Felix Heide.
\newblock Learned feature embeddings for non-line-of-sight imaging and
  recognition.
\newblock {\em ACM Transactions on Graphics (ToG)}, 39(6):1--18, 2020.

\bibitem{dave2020foveated}
Akshat Dave, Muralidhar~Madabhushi Balaji, Prasanna Rangarajan, Ashok
  Veeraraghavan, and Marc~P Christensen.
\newblock Foveated non-line-of-sight imaging.
\newblock In {\em Computational Optical Sensing and Imaging}, pages CTh5C--6.
  Optica Publishing Group, 2020.

\bibitem{dave2022pandora}
Akshat Dave, Yongyi Zhao, and Ashok Veeraraghavan.
\newblock Pandora: Polarization-aided neural decomposition of radiance.
\newblock In {\em Computer Vision--ECCV 2022: 17th European Conference, Tel
  Aviv, Israel, October 23--27, 2022, Proceedings, Part VII}, pages 538--556.
  Springer, 2022.

\bibitem{gariepy2016detection}
Genevieve Gariepy, Francesco Tonolini, Robert Henderson, Jonathan Leach, and
  Daniele Faccio.
\newblock Detection and tracking of moving objects hidden from view.
\newblock {\em Nature Photonics}, 10(1):23--26, 2016.

\bibitem{heide2019non}
Felix Heide, Matthew O’Toole, Kai Zang, David~B Lindell, Steven Diamond, and
  Gordon Wetzstein.
\newblock Non-line-of-sight imaging with partial occluders and surface normals.
\newblock {\em ACM Transactions on Graphics (ToG)}, 38(3):1--10, 2019.

\bibitem{heide2014diffuse}
Felix Heide, Lei Xiao, Wolfgang Heidrich, and Matthias~B Hullin.
\newblock Diffuse mirrors: 3d reconstruction from diffuse indirect illumination
  using inexpensive time-of-flight sensors.
\newblock In {\em Proceedings of the IEEE Conference on Computer Vision and
  Pattern Recognition}, pages 3222--3229, 2014.

\bibitem{henley2022bounce}
Connor Henley, Joseph Hollmann, and Ramesh Raskar.
\newblock Bounce-flash lidar.
\newblock {\em IEEE Transactions on Computational Imaging}, 8:411--424, 2022.

\bibitem{henley2020imaging}
Connor Henley, Tomohiro Maeda, Tristan Swedish, and Ramesh Raskar.
\newblock Imaging behind occluders using two-bounce light.
\newblock In {\em ECCV}, pages 573--588. Springer, 2020.

\bibitem{kadambi2016occluded}
Achuta Kadambi, Hang Zhao, Boxin Shi, and Ramesh Raskar.
\newblock Occluded imaging with time-of-flight sensors.
\newblock {\em ACM Transactions on Graphics (ToG)}, 35(2):1--12, 2016.

\bibitem{katz2014non}
Ori Katz, Pierre Heidmann, Mathias Fink, and Sylvain Gigan.
\newblock Non-invasive single-shot imaging through scattering layers and around
  corners via speckle correlations.
\newblock {\em Nature photonics}, 8(10):784--790, 2014.

\bibitem{kirmani2009looking}
Ahmed Kirmani, Tyler Hutchison, James Davis, and Ramesh Raskar.
\newblock Looking around the corner using transient imaging.
\newblock In {\em 2009 IEEE 12th International Conference on Computer Vision},
  pages 159--166. IEEE, 2009.

\bibitem{kumagai20217}
Oichi Kumagai, Junichi Ohmachi, Masao Matsumura, Shinichiro Yagi, Kenichi Tayu,
  Keitaro Amagawa, Tomohiro Matsukawa, Osamu Ozawa, Daisuke Hirono, Yasuhiro
  Shinozuka, et~al.
\newblock 7.3 a 189$\times$ 600 back-illuminated stacked spad direct
  time-of-flight depth sensor for automotive lidar systems.
\newblock In {\em 2021 IEEE International Solid-State Circuits Conference
  (ISSCC)}, volume~64, pages 110--112. IEEE, 2021.

\bibitem{laurentini1994visual}
Aldo Laurentini.
\newblock The visual hull concept for silhouette-based image understanding.
\newblock {\em PAMI}, 16(2):150--162, 1994.

\bibitem{lin2020passive}
Di Lin, Connor Hashemi, and James~R Leger.
\newblock Passive non-line-of-sight imaging using plenoptic information.
\newblock {\em JOSA A}, 37(4):540--551, 2020.

\bibitem{lindell2019wave}
David~B Lindell, Gordon Wetzstein, and Matthew O'Toole.
\newblock Wave-based non-line-of-sight imaging using fast fk migration.
\newblock {\em ACM Transactions on Graphics (ToG)}, 38(4):1--13, 2019.

\bibitem{liu2019non}
Xiaochun Liu, Ib{\'o}n Guill{\'e}n, Marco La~Manna, Ji~Hyun Nam, Syed~Azer
  Reza, Toan~Huu Le, Adrian Jarabo, Diego Gutierrez, and Andreas Velten.
\newblock Non-line-of-sight imaging using phasor-field virtual wave optics.
\newblock {\em Nature}, 572(7771):620--623, 2019.

\bibitem{maeda2019recent}
Tomohiro Maeda, Guy Satat, Tristan Swedish, Lagnojita Sinha, and Ramesh Raskar.
\newblock Recent advances in imaging around corners.
\newblock {\em arXiv preprint arXiv:1910.05613}, 2019.

\bibitem{metzler2020deep}
Christopher~A Metzler, Felix Heide, Prasana Rangarajan, Muralidhar~Madabhushi
  Balaji, Aparna Viswanath, Ashok Veeraraghavan, and Richard~G Baraniuk.
\newblock Deep-inverse correlography: towards real-time high-resolution
  non-line-of-sight imaging.
\newblock {\em Optica}, 7(1):63--71, 2020.

\bibitem{morimoto2020megapixel}
Kazuhiro Morimoto, Andrei Ardelean, Ming-Lo Wu, Arin~Can Ulku, Ivan~Michel
  Antolovic, Claudio Bruschini, and Edoardo Charbon.
\newblock Megapixel time-gated spad image sensor for 2d and 3d imaging
  applications.
\newblock {\em Optica}, 7(4):346--354, 2020.

\bibitem{o2018confocal}
Matthew O’Toole, David~B Lindell, and Gordon Wetzstein.
\newblock Confocal non-line-of-sight imaging based on the light-cone transform.
\newblock {\em Nature}, 555(7696):338--341, 2018.

\bibitem{pediredla2019snlos}
Adithya Pediredla, Akshat Dave, and Ashok Veeraraghavan.
\newblock Snlos: Non-line-of-sight scanning through temporal focusing.
\newblock In {\em 2019 IEEE International Conference on Computational
  Photography (ICCP)}, pages 1--13. IEEE, 2019.

\bibitem{pediredla2018signal}
Adithya~K Pediredla, Aswin~C Sankaranarayanan, Mauro Buttafava, Alberto Tosi,
  and Ashok Veeraraghavan.
\newblock Signal processing based pile-up compensation for gated single-photon
  avalanche diodes.
\newblock {\em arXiv preprint arXiv:1806.07437}, 2018.

\bibitem{rapp2020seeing}
Joshua Rapp, Charles Saunders, Juli{\'a}n Tachella, John Murray-Bruce, Yoann
  Altmann, Jean-Yves Tourneret, Stephen McLaughlin, Robin Dawson, Franco~NC
  Wong, and Vivek~K Goyal.
\newblock Seeing around corners with edge-resolved transient imaging.
\newblock {\em Nature communications}, 11(1):1--10, 2020.

\bibitem{saunders2019computational}
Charles Saunders, John Murray-Bruce, and Vivek~K Goyal.
\newblock Computational periscopy with an ordinary digital camera.
\newblock {\em Nature}, 565(7740):472--475, 2019.

\bibitem{scheiner2020seeing}
Nicolas Scheiner, Florian Kraus, Fangyin Wei, Buu Phan, Fahim Mannan, Nils
  Appenrodt, Werner Ritter, Jurgen Dickmann, Klaus Dietmayer, Bernhard Sick,
  et~al.
\newblock Seeing around street corners: Non-line-of-sight detection and
  tracking in-the-wild using doppler radar.
\newblock In {\em Proceedings of the IEEE/CVF Conference on Computer Vision and
  Pattern Recognition}, pages 2068--2077, 2020.

\bibitem{sharma2021you}
Prafull Sharma, Miika Aittala, Yoav~Y Schechner, Antonio Torralba, Gregory~W
  Wornell, William~T Freeman, and Fr{\'e}do Durand.
\newblock What you can learn by staring at a blank wall.
\newblock In {\em Proceedings of the IEEE/CVF International Conference on
  Computer Vision}, pages 2330--2339, 2021.

\bibitem{smith2018tracking}
Brandon~M Smith, Matthew O'Toole, and Mohit Gupta.
\newblock Tracking multiple objects outside the line of sight using speckle
  imaging.
\newblock In {\em Proceedings of the IEEE Conference on Computer Vision and
  Pattern Recognition}, pages 6258--6266, 2018.

\bibitem{swedish2021objects}
Tristan Swedish, Connor Henley, and Ramesh Raskar.
\newblock Objects as cameras: Estimating high-frequency illumination from
  shadows.
\newblock In {\em Proceedings of the IEEE/CVF International Conference on
  Computer Vision}, pages 2593--2602, 2021.

\bibitem{tanaka2020polarized}
Kenichiro Tanaka, Yasuhiro Mukaigawa, and Achuta Kadambi.
\newblock Polarized non-line-of-sight imaging.
\newblock In {\em Proceedings of the IEEE/CVF Conference on Computer Vision and
  Pattern Recognition}, pages 2136--2145, 2020.

\bibitem{tancik2018flash}
Matthew Tancik, Guy Satat, and Ramesh Raskar.
\newblock Flash photography for data-driven hidden scene recovery.
\newblock {\em arXiv preprint arXiv:1810.11710}, 2018.

\bibitem{tiwary2022orca}
Kushagra Tiwary, Askhat Dave, Nikhil Behari, Tzofi Klinghoffer, Ashok
  Veeraraghavan, and Ramesh Raskar.
\newblock Orca: Glossy objects as radiance field cameras.
\newblock {\em arXiv preprint arXiv:2212.04531}, 2022.

\bibitem{velten2012recovering}
Andreas Velten, Thomas Willwacher, Otkrist Gupta, Ashok Veeraraghavan, Moungi~G
  Bawendi, and Ramesh Raskar.
\newblock Recovering three-dimensional shape around a corner using ultrafast
  time-of-flight imaging.
\newblock {\em Nature communications}, 3(1):1--8, 2012.

\bibitem{willomitzer2019high}
Florian Willomitzer, Fengqiang Li, Muralidhar~Madabhushi Balaji, Prasanna
  Rangarajan, and Oliver Cossairt.
\newblock High resolution non-line-of-sight imaging with superheterodyne remote
  digital holography.
\newblock In {\em Computational Optical Sensing and Imaging}, pages CM2A--2.
  Optica Publishing Group, 2019.

\bibitem{willomitzer2021fast}
Florian Willomitzer, Prasanna~V Rangarajan, Fengqiang Li, Muralidhar~M Balaji,
  Marc~P Christensen, and Oliver Cossairt.
\newblock Fast non-line-of-sight imaging with high-resolution and wide field of
  view using synthetic wavelength holography.
\newblock {\em Nature communications}, 12(1):6647, 2021.

\bibitem{xin2019theory}
Shumian Xin, Sotiris Nousias, Kiriakos~N Kutulakos, Aswin~C Sankaranarayanan,
  Srinivasa~G Narasimhan, and Ioannis Gkioulekas.
\newblock A theory of fermat paths for non-line-of-sight shape reconstruction.
\newblock In {\em Proceedings of the IEEE/CVF Conference on Computer Vision and
  Pattern Recognition}, pages 6800--6809, 2019.

\bibitem{yue2022cornerradar}
Shichao Yue, Hao He, Peng Cao, Kaiwen Zha, Masayuki Koizumi, and Dina Katabi.
\newblock Cornerradar: Rf-based indoor localization around corners.
\newblock {\em Proceedings of the ACM on Interactive, Mobile, Wearable and
  Ubiquitous Technologies}, 6(1):1--24, 2022.

\bibitem{zhang2021240}
Chao Zhang, Ning Zhang, Zhijie Ma, Letian Wang, Yu Qin, Jieyang Jia, and Kai
  Zang.
\newblock A 240$\times$ 160 3d-stacked spad dtof image sensor with rolling
  shutter and in-pixel histogram for mobile devices.
\newblock {\em IEEE Open Journal of the Solid-State Circuits Society}, 2:3--11,
  2021.

\end{thebibliography}
}

\end{document}